\documentclass[12pt,twocolumn]{article}
\usepackage{hyperref} 
\hypersetup{
    bookmarks=true,         % show bookmarks bar?
    unicode=false,          % non-Latin characters in Acrobat’s bookmarks
    pdftoolbar=true,        % show Acrobat’s toolbar?
    pdfmenubar=true,        % show Acrobat’s menu?
    pdffitwindow=false,     % window fit to page when opened
    pdfstartview={FitH},    % fits the width of the page to the window
    pdftitle={MASER Navigation},    % title
    pdfauthor={Dong Jiang},     % author
    pdfsubject={Radio Celestial Navigation},   % subject of the document
    pdfcreator={DJ},   % creator of the document
    pdfproducer={DJ}, % producer of the document
    pdfkeywords={Radio Navigation MASER}, % list of keywords
    pdfnewwindow=true,      % links in new window
    colorlinks=true,       % false: boxed links; true: colored links
    linkcolor=red,          % color of internal links
    citecolor=green,        % color of links to bibliography
    filecolor=magenta,      % color of file links
    urlcolor=cyan           % color of external links
}

\usepackage{graphicx}
\usepackage{times}
\usepackage{natbib}

\usepackage{aas_macros}
\usepackage{amsmath,amssymb}
\newcommand{\degr}{\ensuremath{^{\circ}}}

\begin{document} 

\title{The Principle and Application of Maser Navigation}
 
\author{Jiang Dong,%
\thanks{DJ is in Yun Nan Astronomical Observatory, E-mail: \href{mailto:djcosmic@gmail.com}{djcosmic@gmail.com},
and Scientific and application center of lunar and deep space exploration, NAOs, China. The content of this paper have applied the patent.}
\thanks{Manuscript received ---; revised ---.}}

\maketitle

\begin{abstract}
The traditional celestial navigation system (CNS) is used 
the moon, stars, and planets as celestial guides. 
Then the star tracker (i.e. track one star or planet or angle between it)
and star sensor (i.e. sense many star simultaneous) 
be used to determine the attitude of the spacecraft. 
Pulsar navigation also be introduced to CNS. 
Maser is another interested celestial in radio astronomy 
which has strong flux density as spectral line. 
Now we analysis the principle of maser navigation which 
base on measuring Doppler shift frequency spectra  
and the feasibility that use the exist instrument.
We give the navigation equations of maser-based navigation system
and discuss the integrated navigation use maser,
then give the perspective in the Milky Way and the intergalatic. 
Our analysis show that use one meter antenna can achieve tens of meters 
position accuracy which better than today's star sensor.  
After integrated with maser navigation, pulsar navigation and star sensor 
in CNS and inertial navigation system, 
is it not only increase the reliability and redundancy of 
navigation or guiding system 
but also can less or abolish the depend of Global Navigation
Satellite System (GNSS) which include GPS, GRONSS, Galileo and BeiDou et al. 
Maser navigation can give the continuous position in deep space,
that means we can freedom fly successfully in the Milky Way 
which use celestial navigation that
include maser, pulsar and traditional star sensor.
Maser as nature beacon in the universe will 
make human freely fly in the space of the Milky Way, even outer of it. 
That is extraordinary in the human evolution to type III of Kardashev
civilizations.

\end{abstract}

%\keywords{masers -- [celestial navigation]}

\section{Introduction}
Maser is a device that produces coherent electromagnetic waves 
through amplification due to stimulated emission. 
Historically the term came from the acronym 
``Microwave Amplification by Stimulated Emission of Radiation'', 
although modern masers emit 
over a broad portion of the electromagnetic spectrum. 
Astrophysical maser is a naturally occurring source
of stimulated spectral line emission, typically
in the microwave portion of the electromagnetic spectrum. 
It was discovered by Weaver, H. et al. firstly \citep{wwdl65}, 
after Charles Townes given prediction and Weinreb, S. et al. 
firstly detected the hydroxyl molecule (OH) \citep{wbmh63}, 
that was the first radio observation of an interstellar molecule. 
This emission may arise in molecular clouds, comets, planetary atmospheres, 
stellar atmospheres, or from various conditions in interstellar space.

The traditional CNS origin from nautical, 
developed to aeronautics by US (B52, B-1B, B-2A, C-141A, SR-71, F22 et al.)
and Soviet (TU-16, TU-95, TU-160 et al.) \citep{pdl+01,w07}, 
success in determine the attitude of the spacecraft in
help orient the Apollo spacecraft enroute to and from the Moon. 
Although the GNSS and Inertial Navigation System (INS) 
almost can finish any job in this planet now, 
someone still continued think it is important for 
it can be used independently of ground aids and has global coverage, 
it cannot be jammed (except by clouds) and 
does not give off any signals that could be detected by an the others. 
The traditional maritime state which include US, Russia, UK, French, 
all spend many money in CNS for its unique advantage.

Pulsar navigation is use pulsar 
as beacon give the continuous position in deep space.
Dr Sheikh et al. construct the X-ray pulsar-based autonomous navigation theory  
which based modern spacecraft navigation technique
that include Kalman filter et al. \citep{s05,s07}. 
Dong Jiang analysis the feasibility that use radio pulsar navigation and
discuss the integrated navigation use pulsar,
then give the different navigation mission analysis and 
design process basically which include 
the space, the airborne, the ship and the land of the planet 
or the lunar in the solar system \citep{dj08a}. 
With the distance increase, 
the radiometric tracking of deep space network (DSN) will decrease in accuracy, 
and it can not work when spacecraft in the other side of sun \citep{rsg+08} 
and land rover in the back of the other planet or lunar. 
But pulsar can not be effected in that place.

A maser-based navigational system is considered by Shapiro et al. 
using the emissions from ${\rm H_2O}$ molecules which are the most intense 
in Very Long Baseline Interferometry (VLBI) navigation \citep{suy72}. 
The VLBI technique, with a master station, 
can use either an artificial satellite 
or natural sources as position references, a high-speed data link is required. 
The characteristics of natural radio sources, 
their flux, distribution on the sky, and apparent size are shown to 
provide a limit on position measurement precision \citep{kjh73}. 
Then Wallace, K. discuss that use radio sextant and radio star 
which include maser to navigation \citep{w88}
that just is the geometry method of traditional nautical celestial navigation.
They thought the accuracy of a ``radio sextant'' is dependent amongst 
other things on the signal bandwidth, and the line emission 
which is typically in the region of 50 kHz is 
too narrow to attain reasonable fix accuracies ($< 10\ {\rm nm}$)
without the use of long-baseline interferometric techniques.

Now I analysis the principle of maser autonomous navigation system 
which base measure Doppler shift frequency spectra in one receiver
that is the similar process of measure Doppler effect in radio navigation,
astrodynamic and spacecraft navigation technique.

\section{Principle of Maser Navigation}
\subsection{Doppler effect in Maser Observation}
\begin{figure}
\includegraphics[width=7cm,height=10cm]{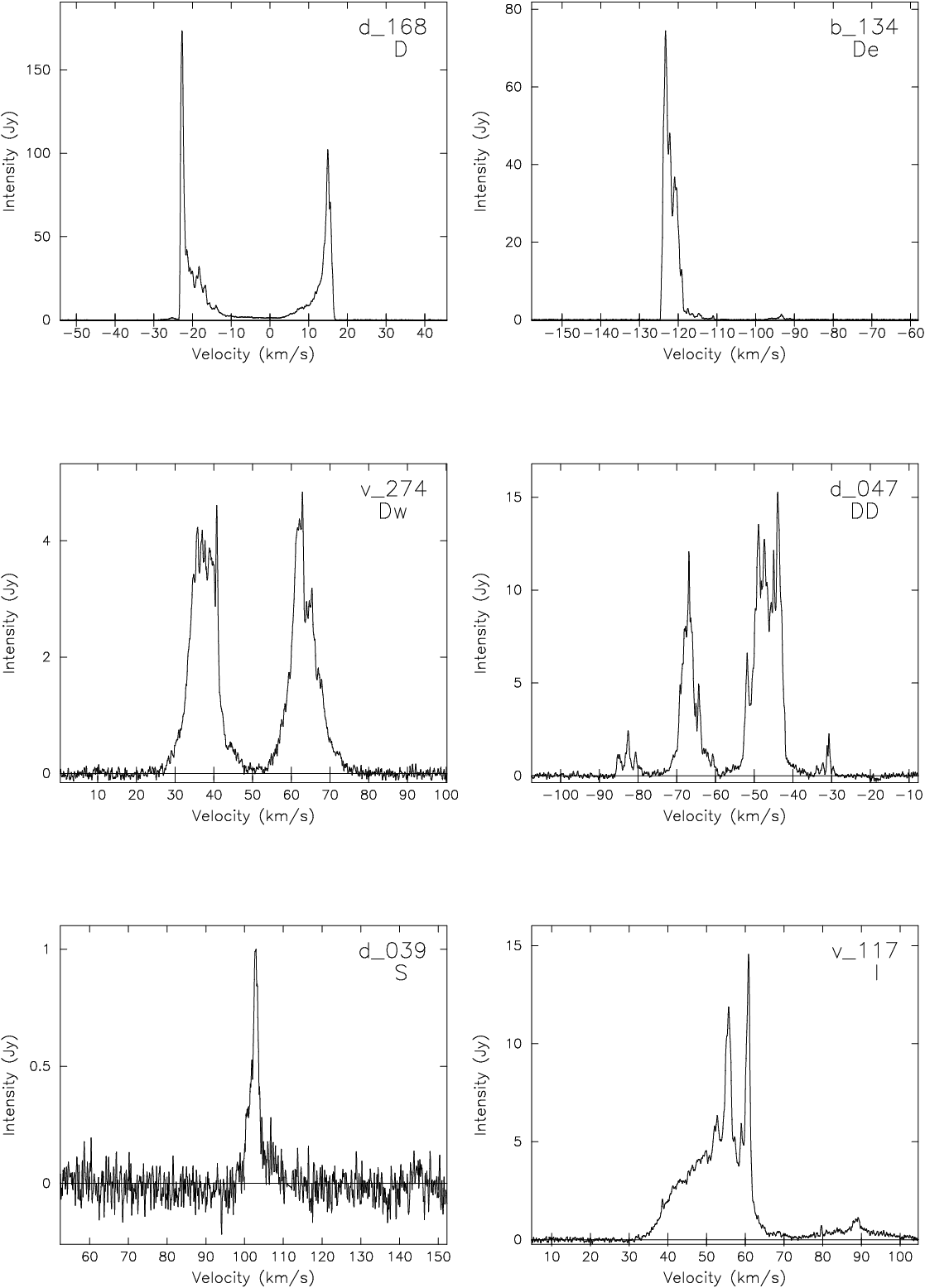}
\vspace{-1.5cm}
\caption[]{}
\label{fig:profile}
\end{figure}

Maser emission from molecules such as water, hydroxyl (OH), 
and silicon monoxide (SiO) is strong spectral line that 
is an important tracer of the gas kinematics and magnetic field
strength in astrophysical interesting regions. 
Figure. \ref{fig:profile} show some examples of spectra from maser 
in Post-AGB stars \citep{dgc04}.
The order of velocity is dozens of kilometers per second in this figure.
Some of it have double peaks structure for the
Doppler effect that come from the rotation of star.

The Doppler effect (or Doppler shift) is the change in frequency and wavelength
of a wave for an observer moving relative to the source of the waves.
It is commonly heard when a vehicle sounding a siren approaches, 
passes and recedes from an observer. 
The received frequency is increased (compared to the emitted frequency) 
during the approach, it is identical at the instant of passing by, 
and it is decreased during the recession.
For waves which do not require a medium, such as light or gravity in
special relativity, only the relative difference in velocity 
between the observer and the source needs to be considered.
The Doppler effect for electromagnetic waves such as light is of great
use in astronomy and results in either a so-called redshift or blueshift. 
It has been used to measure the speed at which stars and galaxies 
are approaching or receding from us, that is, the radial velocity. 
This is used to detect if an apparently single star is, in reality, 
a close binary and even to measure the rotational speed of stars and galaxies.

According to the relativistic Doppler effect, we will have the relation
between the frequency we will receive $f^{'} $ and 
the frequency the source emission $f_{0} $:
$$
{f^{'} \over f_{0}} = {{\sqrt {1-\beta^2}} \over {1-\beta*cos\theta}} ~,
$$
here $\beta= {\nu \over c}$, 
$\nu $ is the relative velocity between the source and the observer, 
$\theta $ is the angle between the line of the source to the observer
and the direction of the source movement, c is the speed of light.

When $ {\theta} = {90 \degr} $, called transverse Doppler effect, 
we have the relation:
$$
{f^{'} \over  f_{0}} = {\sqrt {1-\beta^2}} ~.
$$
When $\theta = 0\degr $ and $\theta = 180\degr $, 
called longitudinal Doppler effect, we have the relation:
$$
{f^{'} \over  f_{0}} = {\sqrt {{1+\beta}\over {1-\beta}}} ~,
$$ 
and  
$$
{f^{'} \over  f_{0}} = {\sqrt {{1-\beta}\over {1+\beta}}} ~. 
$$
In usual, the transverse Doppler effect far less than 
longitudinal Doppler effect, so astronomer only 
calculate longitudinal Doppler effect.
When $\nu \ll c$, we have:
$$
{f^{'} \over  f_{0}} = {1^{+}_{-}\beta} ~.
$$
So the value of Doppler shift spectra  
\begin{equation}
 \Delta f = {f^{'} -  f_{0}} = {^{+}_{-} f_{0}\beta} ~.
\label{eq:doppler}
\end{equation}
In the formula, ``+'', ``-'' correspond to blueshift and redshift.

\begin{figure}
\includegraphics[width=5cm,height=5cm]{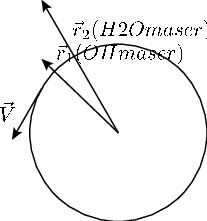}
\vspace{-1.5cm}
\caption[]{}
\label{fig:2D}
\end{figure}

%%Doppler effect make line widen $${f^{'}_{c}} = {f_{0}(1^{+}_{-}\beta)} $$
From the above formula, if we know the frequency $f^{'} $ and 
the frequency the source emission $f_{0} $,
we will have the relative velocity between the source and the observer.
In astronomical observation, the velocity be normalized to 
the local standard of rest (LSR) for it benefit to study the celestial 
in a uniform frame.
So I think we can use the Doppler effect to navigation.
Figure. \ref{fig:2D} show the principle of maser navigation in two dimension.
The center is LSR, V is the velocity of spacecraft.
If we can receive two signal which come from maser sources,
we will have the relative velocity 
between the observer (i.e. the vehicle) and the LSR. 
Then using the velocity plus the time, we will have 
the relative position  between the vehicle and the LSR.
The similar principle of maser navigation in three dimension, 
we will have the information of the continuous position in the space,
if we can receive three maser signal simultaneous.

 \subsection{Kalman filter for Maser Navigation}
The kalman filter is an efficient recursive linear filter 
that estimates the state of a dynamic system 
from a series of noisy measurements \citep{k60}.
It is mainly used to estimate system states 
that can only be observed indirectly or inaccurately by the system itself.
It can predict the motion of anything for it is recursive,
even the signal have noise for that use the dynamic state estimate the system. 
In maser navigation, that is significant like it in INS and 
the traditional CNS (i.e. star sensor).
We can use navigation kalman filter measure
spectral line range, spacecraft clock,
then compare with the signal which come from maser,  
so we will have the velocity and position through plus time.

In order to use the kalman filter, 
one must model the process in accordance with 
the framework of the kalman filter. 
This means specifying the following matrices: 
$\mathbf{A}_k$, the state-transition model; 
$\mathbf{H}_k$, the observation model; 
$\mathbf{Q}_k$, the covariance of the process noise; 
$\mathbf{R}_k$, the covariance of the observation noise; 
and sometimes $\mathbf{B}_k$, the control-input model, 
for each time-step, $k$, as described below.
The kalman filter model 
assumes the true state at time $k$ is evolved 
from the state at $(k-1)$ according to
\begin{equation}
\textbf{x}_{k} = \textbf{A}_{k}\textbf{x}_{k-1} + 
\textbf{B}_{k}\textbf{u}_{k} + \textbf{w}_{k} ~,
\label{eq:xevol}
\end{equation}
where$\mathbf{A}_k$ is applied to the previous state $\mathbf{x}_{k−1}$;
$\mathbf{B}_k$ is applied to the control vector $\mathbf{u}_k$;
$\mathbf{w}_k$ is the process noise which is assumed to be drawn from 
a zero mean multivariate normal distribution with covariance $\mathbf{Q}_k$,
$\mathbf{w}_{k} \sim \mathbf{N}(0, \mathbf{Q}_k) $.
At time k an observation (or measurement) $\mathbf{z}_k$ of 
the true state $\mathbf{x}_k$ is made according to
\begin{equation}
    \textbf{z}_{k} = \textbf{H}_{k} \textbf{x}_{k} + \textbf{v}_{k}~,
\label{eq:zofx}
\end{equation}
where $\mathbf{H}_k$ maps the true state space into the observed space and 
$\mathbf{v}_k$ is the observation noise which is assumed to 
be zero mean Gaussian white noise with covariance $\mathbf{R}_k$,
$\mathbf{v}_{k} \sim \mathbf{N}(0, \mathbf{R}_k)$.
The initial state, and the noise vectors at each step 
${x_0, w_1, ..., w_k, v_1 ... v_k}$ are all assumed to be mutually independent.

\begin{figure}
\includegraphics[width=7cm,height=6cm]{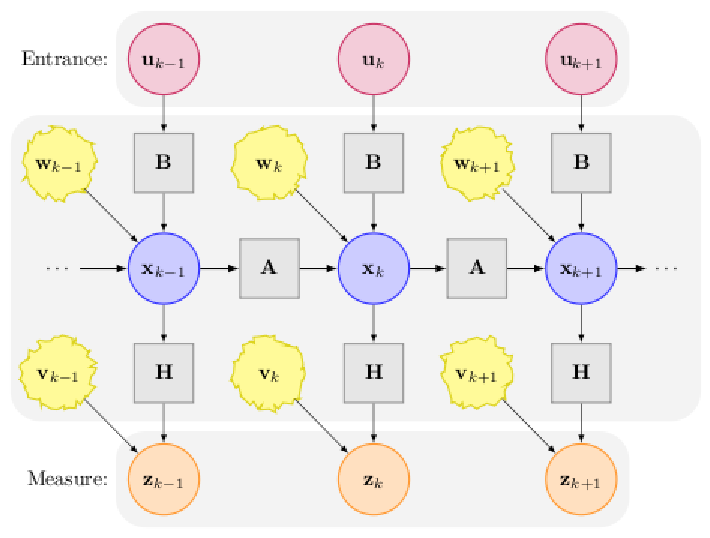}
\vspace{-1.5cm}
\caption[]{}
\label{fig:kf}
\end{figure}

Figure \ref{fig:kf} show the system model of the (linear) kalman filter.
At each time step the state vector $\mathbf{x}_k$ is propagated 
to the new state estimation $\mathbf{x}_{k+1}$ 
by multiplication with the constant state transition matrix $\mathbf{A}$. 
The state vector $\mathbf{x}_{k+1}$ is additionally influenced 
by the control input vector $\mathbf{u}_{k+1}$ multiplied 
by the input matrix $\mathbf{B}$, 
and the system noise vector $\mathbf{w}_{k+1}$. 
The system state cannot be measured directly. 
The measurement vector $\mathbf{z}_k$ consists of the information contained
within the state vector $\mathbf{x}_k$ multiplied by the measurement
matrix $\mathbf{H}$, and the additional measurement noise $\mathbf{v}_k$.

\subsection{Navigation Equations of Maser-based Navigation System}
The receiver uses messages received from three masers to determine 
the telescope positions. The x, y, and z components of velocity 
sent are designated as $[V_x, V_y, V_z]$ ,
where the subscript $i$ denotes the masers and has the value 1, 2 or 3.
So we have navigation equations: 
\begin{equation}
\vec{V} = V_x^2 + V_y^2 + V_z^2 ~.
\label{eq:ne}
\end{equation}
Here, the direct measurements is velocity which relative to masers.

We will have the instantaneous change of the position that 
use the instantaneous velocity multiplied with time, 
the position equation is : 
\begin{equation}
\delta S = \delta \vec{V} \times \delta t ~.
\label{eq:nes}
\end{equation}

If we only can observe two or one maser,
we still can perform maser navigation in combination with 
the orbit of the mobile station that use kalman filter et al.
If it is two masers, we can give the weighted mean of position and time 
from the equation:
\begin{equation}
x^* =  \frac{\sum p_i x_i}{\sum p_i} ~.
\label{eq:nem}
\end{equation}
Here the weight of the measurement $x_i$ is $p_i$,
$p_i = \sigma_1^2 / \sigma_i^2$ $(i = 1 \sim N)$, 
$\sigma_i$ the standard deviation of unequal precision measurement sequence,
we set $\sigma_1 = max \sigma_i$ ($i = 1 \sim N$) in usual.

\section{Maser signal process in astronomy Vs 
The requires of engineer  project Vs The reliable of technique}
Maser have been found in transitions of OH, SiO,
water, methanol, ammonia, and other molecules, 
and also in recombination lines of hydrogen.
The maser observation system sensitivity (i.e. the raw limiting flux density) 
is given by the radiometer equation:
\begin{equation}
S_{\rm lim} = \frac{\sigma \beta}{(B N_{\rm p} \tau_{obs})^{1/2}}
{\frac{T_{\rm sys}}{G}} ~,
\label{eq:s}
\end{equation}
where $\sigma$ is a loss factor, taken to be 1.5 (One-bit sampling 
at the Nyquist rate introduces a loss of $\sqrt{2/\pi}$ 
relative to a fully sampled signal. 
The principal remaining loss results from 
the non-rectangular bandpass of the channel filters).  
$\beta$ is the detection signal-to-noise ratio threshold, taken to be 5.0, 
$G$ is the telescope gain, $B$ is the receiver bandwidth in Hz, 
$N_{\rm p}$ is the number of polarizations
and $\tau_{\rm obs}$ is the time per observation in seconds.
$T_{\rm sys}$ is the system temperature, $G$ is the telescope gain, 
$G = A_{\rm e}/(2k_{\rm B})$,
here $A_{\rm e}$ is the effective area of a telescope, 
$k_{\rm B}$ is Boltzmann's constant.

Using the above formula, we use 4 M antenna 
(If the telescope efficiency is 0.8,
$A_{\rm e} = 0.8 \times \pi(4/2)^2 \simeq 10\ {\rm m^2}$, 
$G \simeq 3.62 \times 10^{-3}\ {\rm KJy^{-1}}$),   
we set $T_{\rm sys}$ is 20 K, $B$ is 1 MHz , $N_{\rm p}$ is 2, 
$\tau_{\rm obs}$ is 4 min,
$\lambda_{0}$ is 22.2 GHz (${\rm H_2O}$ Interstellar Maser),
so we have $S_{\rm lim} \simeq 1.89\ {\rm Jy}$.
If we set $B$ is 10 kHz, $\tau_{\rm obs}$ is 1 sec, the other is not change,
we have $S_{\rm lim} \simeq 293\ {\rm Jy}$.
The flux density of maser is very biggest,
The table ~\ref{Sources} is the type list of the strong radio maser source,
we can observed it even if use 4 meter antenna in microwave in one second.
If the diameter of dish is 1 M (If the telescope efficiency is 0.8,
$A_{\rm e} = 0.8 \times \pi(1/2)^2 \simeq 0.68 \times 10^{-1}\ {\rm m^2}$, 
$G \simeq 2.27 \times 10^{-4}\ {\rm KJy^{-1}}$),   
we set $T_{\rm sys}$ is 20 K, $B$ is 10 kHz , $N_{\rm p}$ is 2, 
$\tau_{\rm obs}$ is 4 sec,
$\lambda_{0}$ is 22.2 GHz (${\rm H_2O}$ Interstellar Maser),
so we have $S_{\rm lim} \simeq 2.34 \times 10^3\ {\rm Jy}$.
From the table  ~\ref{Sources}, 
we Still can use ${\rm H_2O}$ maser to navigation.

The above formula use Jy as unit.
The Jansky (Jy) is a measure of spectral power flux density - 
the amount of RF energy per unit time per unit area per unit bandwidth,
$1\ {\rm Jy} \equiv 10^{-26}\ {\rm W/m^2/Hz}$.
The jansky is not used outside of radio astronomy.
It is not a practical unit for measuring communications signals,
the magnitude is much too small, and is a linear unit,
very few RF engineers outside of radio astronomy will know what a Jy is.
Because of wide dynamic range encountered the most radio systems,
the power is usually expressed in logarithmic units of 
watts (dBW) or milliwatts (dBm):
${\rm dBW} \equiv 10log_{10}Power_{\rm watts}$,
${\rm dBm} \equiv 10log_{10}Power_{\rm milliwatts}$.
While not comprised of the same units,
we can make some reasonable assumptions to compare a Jy to dBm.
Assumptions bandwidth is 1 MHz, 22.2 GHz frequency 
($\lambda_{0} =0.01\ {\rm m}$), parabolic receive antenna, 
antenna collecting area $= \pi \times r^2 = 3.14 \times (4/2)^2 
= 12.6\ {\rm m^2}$. How much is one Jy worth in dBm ?
$P_{\rm mW}= 10^{-26}\ {\rm W/m^2/Hz} \times 1,000,000\ {\rm Hz} 
\times 12.6\ {\rm m^2} \times 1000\ {\rm mW/W} 
= 1.26 \times 10^{-16}\ {\rm mW}$,
$P_{\rm dBm} = 10log(1.26\times 10^{-16}\ {\rm mW}) = -158.9963\ {\rm dBm}$.
Considering the parabolic antenna as a circular aperture 
gives the following approximation for the maximum gain: 
$ G_{\rm dBi} \simeq 10log((9.87 \times D^2)/\lambda_{0}^2$.
in this form, $G$ is power gain over isotropic
$D$ is reflector diameter in same units as wavelength, 
$\lambda_{0} $is center of wavelength.
For 4 M diameter and $\lambda_{0} =0.013\ {\rm m}$, $ G_{\rm dBi} =59.3777$.
So 1 Jy in 4 M antenna is $-99.6186\ {\rm dBm}$.
If we set $10^3\ {\rm Jy}$, we have $P_{\rm dBm} = −128.9963\ {\rm dBm}$, 
after 4 M antenna amplify it, we have the signal is $−69.6186\ {\rm dBm}$.

Masers take place in several places in the universe: 
in the vicinity of newly forming stars and 
regions of ionized hydrogen (OH, water, SiO, and methanol masers); 
in the circumstellar shells of stars at the end of its life — that is, 
red giants and supergiants (OH, water, and SiO masers); 
in the shocked regions where supernova remnants are expanding 
into an adjacent molecular cloud (OH masers);
and in the nuclei and jets of active galaxies (OH and water masers).
The emission from OH masers can vary on timescales of hundreds of seconds 
and be detected as long-duration radio bursts \citep{cb85,y86}.
In the above, maser from circumstellar matter of red giants and supergiants 
(i.e. AGB and post-AGB) is well in navigation for 
some of it have double peaks structure that easily identified.
The emission from maser of circumstellar matter have vary on 
timescales of orders of three months to years \citep{ebsl01, lmrt01}.

Navigation of use maser just for a continuous spectral line signal 
during the different mission time which during tens of minutes to several years.
When we penetrate the system of maser navigation as one systems engineering,
I think navigation system use maser is feasible absolutely.
In maser navigation, radial velocity measurements 
and the time measurements is important to have the position.
The accuracy of this navigation system only depends on 
the accuracy of the spectrum we have obtained.
From equation \ref{eq:s}, we have $ V = \Delta f \times c / f_{0}$,
so 1 kHz shift in 22.2 GHz corresponds to about $13.5\ {\rm ms^{-1}}$.
Today these spectrometers (autocorrelators, acusto-optical spectrometers 
and filterbank) offer a useable bandwidth from a few kHz up to 2 GHz 
with a few thousand spectral channels, 
which are capable of resolving narrow spectral lines of masers.
If we can measure 1 Hz shift of ${\rm H_2O}$ maser, 
we will have the velocity accuracy is $1.35\ {\rm cms^{-1}}$.  
The wide bandwidth of receiver will give the big measuring range of velocity.
It also is important to maser navigation.

The light frequency comb have developed in recently \citep{hall06,h06}
that will play key role in maser navigation. 
A laser frequency comb that enables radial velocity measurements 
with a precision of $1\ {\rm cms^{-1}}$ \citep{lbf+08}.
If we can achieve the similar instrument in microwave,
we can have easily finish maser navigation.
The atomic clock has the advantage that keep time in short timescale.
Pulsar especially millisecond pulsars (MSP) be thought 
the nature’s most stable clocks \citep{t91}. 
The data show some pulsar stability than atomic clock 
in timescale than one year \citep{mte97}.
When integrated it, even plus light frequency comb clock 
in the future, that will satisfied with maser navigation.   
Some modern digital signal processing (DSP) technique can 
apply to maser signal navigation which include weak signal detection,
signal enhancement, signal reconstruct et al.
Maser spectral line is Gaussian for the thermal motion of molecule,
but the complex surrounding for example turbulence 
make profile become complex.  
In navigation, we just need the information from phase,
so we can magnify the weak maser profile signal 
through normalizing it to a Gaussian signal or plus a Gaussian signal.
The navigation system must leave a copy of raw data to astronomer
for the best filter is construct a good noise model by it.

Dong Jiang analysis the special parabolic dish use in spacecraft
to achieve pulsar tracker, the phased array antenna to achieve pulsar sensor,
and the phased array feed can apply in pulsar sensor 
when use one dish \citep{dj08a}.
The similar technique can use to maser sensor in navigation.
The phased array antenna or feed can receive 
several radio celestial which include different maser or pulsar simultaneous.   
With electronic technique development,
high speed analog-to-digital converter (ADC) 
obtain order of ${\rm Gigabyte^{-s}}$ easily,
field-programmable gate array (FPGA),
multi-core multi-PC cluster and graphics processing unit (GPU)
all apply to scientific computing.
If we can fuse Multi-core CPU, GPU and FPGA, 
construct one computing server and use different advantage of it,
That will easily finish many scientific computation which 
include reduce different radio sources.

Integrated navigation with maser between pulsar navigation, 
CNS and INS, even GNSS,
is realistic path in the future mission.
It will increase the reliability and redundancy of 
navigation or guiding system \citep{zz08}.
The multi-waveband maser navigation also is interested.
In integrated navigation, system analysis and modeling,
system state estimation, filter design, 
information synchronization and system fault tolerance filter design
all is important.
Dong Jiang give the different navigation mission analysis and design
process basically which include the space, the airborne, the ship and 
the land of the planet or the lunar in the solar system \citep{dj08a}.
The similar analysis also fit for maser navigation.

\section{Maser Navigation In the Milky Way and Intergalaxy}
The virtue is obvious, when the rover in the back of   
the others planet or lunar, DSN can not work and 
human can not built GNSS for the other planet in long term. 
So the maser navigation and radio pulsar navigation is one and only method
at any place of the other planet surface day and night in the future explore. 
The advantage of maser navigation is some of beacon that maser emission 
come from the nuclei and jets of active galaxies 
(OH and water masers) \citep{l2005}.
So human will have chance use it freely fly 
in the space of the Milky Way and Intergalaxy.  

Soviet astronomer Kardashev N. S. proposed 
a scheme for classifying advanced technological civilizations.
He identified three possible types and distinguished 
between them in terms of the power 
they could muster for the purposes of interstellar communications. 
The Type III civilization would have evolved far enough to 
tap the energy resources of an entire galaxy. 
This would give a further increase by at least a factor of 10 billion 
to about $10^{36}$ watts \citep{k64}. 
If we want to use the resources of someplace,
we must freely voyage in the that space firstly.
Maser navigation must be extraordinary in the human evolution to 
type III of Kardashev civilizations.

Now the star sensor in optic can give the position accuracy is 
100 m ($3\sigma$). Digital spectrum analyzer can provide the resolution
is 100 Hz in frequency domain in today,
so it can provide the accuracy about $1.35\ {\rm ms^{-1}}$ in velocity and
the position accuracy is several meters.
This means that maser navigation is better in the position accuracy than
star sensor which does not require any technological breakthroughs. 
In order to get better result from maser navigation system,
we still need the biggest telescope find the more masers,
and the special telescope study maser's noise and astrometric model, 
and the corresponding software and hardware to improve 
the spectral lines shift estimater of maser.

\section*{Acknowledgments}
DJ thanks Boffin Chen Pei-Sheng advise that 
OH maser of CSE is well in navigation.

\bibliographystyle{apsr}
\bibliography{mnbib}

\clearpage

%% TABLES
\begin{table*}[htbp]
\caption{The classical parameter of the strong Masers. \label{Sources}}
\begin{center}
\begin{tabular}{l|c|c|c|c}
\hline
Name                    & F         & $T_{\rm B}$ & $\Delta\nu/\delta\nu $ & \\
Maser class             & (Jy)      & (K)   &(${\rm km*s^{-1}/km*s^{-1}}$) & \\
\hline
OH-Stellar Maser (1612 MHz)       & $2 \times 10^2$ & $10^8$   & $1/30$ & \\  
$\rm H_2O$-Stellar Maser (22.2 GHz) & $4 \times 10^3$ & $10^{11}$ & $1/20$ & \\
SiO-Stellar Maser (43.1 GHz)      & $2 \times 10^2$ & $10^{10}$ & $1/10$ & \\
OH-Interstellar Maser (1665 MHz)  & $2 \times 10^2$ & $10^{12}$ & $0.1/50$ & \\
$\rm H_2O$-Interstellar Maser (22.2 GHz) & $4 \times 10^3$ & $10^{14}$ & $1/50$ & \\
$\rm CH_3OH$-Interstellar Maser (12.1 GHz)& $5 \times 10^2$ & $>10^{12}$ & $1/5$ & \\
\hline
\end{tabular}
\end{center}
F is flux density, $T_{\rm B}$ is the brightness temperature, 
$\Delta\nu/\delta\nu $ is the ratio of the velocity width of maser and  
the velocity range of the whole maser.
\end{table*}

%%figure
\noindent {\bf Fig. 1.} Some examples of spectra from maser in 
Post-AGB Stars\citep{dgc04}.

\noindent {\bf Fig. 2.} The principle of maser navigation in two dimension.

\noindent {\bf Fig. 3.} This is the system model of the (linear) kalman filter.
The picture download from net, author is Burkart Lingner.

\end{document}